\documentclass[11pt]{article}
\usepackage{times}
\usepackage{graphicx}

\begin{document}
\title{Relations between the modified Chaplygin gas and a scalar field}
\author{Sandro Silva e Costa\\
{\normalsize \it Centro de Ci\^{e}ncias Naturais e Humanas, Universidade Federal do ABC} \\
{\normalsize \it Rua Santa Ad\'{e}lia, 166 --  
Santo Andr\'{e} -- SP, 09210-170, Brazil} \\
{\it e-mail}: sandro.costa@ufabc.edu.br}
\date{\today}
\maketitle

\begin{abstract}
In this work the modified Chaplygin gas is related to a cosmological scalar field. Analytical results, more general than the solutions previously shown in the literature, are presented for the case when the curvature is absent, and one entirely new particular result is shown for a case where there is curvature. Also, it is emphasized here that one interesting feature of the scalar field associated to the modified Chaplygin gas is the possible presence of two minima in its potential. Finally, it is argued that the scalar field representation of the Chaplygin gas can be another useful tool in the study of the evolution of perturbations and, as an example of this, a new analytical result for the density contrast is presented.
\end{abstract}
\section{Introduction}
Among the several ideas being discussed presently in cosmology, one can find the interesting proposal of unifying dark matter and dark energy with the use of a single component with an 'exotic' equation of state \cite{linear}-\cite{Flavia}. A somewhat popular candidate for playing the role of a unified dark matter is the so-called Chaplygin gas, an exotic fluid whose main characteristic is to have the product between its pressure $p$ and its energy density $\rho$ as a negative constant, i.e.\footnote{Throughout the entire text natural units are used, i.e., $c=G=\hbar=k_B=1$.}, $p\rho=-M$. 

Many works can be found in the literature studying the implications of the use of the Chaplygin gas and its generalizations as a cosmological fluid\footnote{A recent proposal by Debnath \cite{Debnath2}, christened as variable modified Chaplygin gas, seems to be the ultimate generalization, reproducing, for example, the result obtained by de Berredo-Peixoto, Shapiro and Sobreira \cite{Flavia} from a completely different approach.}. Some works indicate how to relate the several Chaplygin gas models to a simple homogeneous scalar field \cite{alternative}-\cite{Perrotta}, or to a more sophisticated Born-Infeld-like Lagrangian \cite{BornInfeld1}. An interesting recent study \cite{linear} relates a generalization of the Chaplygin gas to a scalar field in order to adress the issue of how to distinguish between the generalized Chaplygin gas and the standard cosmological model ($\Lambda$-CDM), and there it is argued that ``implementing the generalized Chaplygin gas as a scalar field obeying a particular Lagrangian is useful on a number of levels.''  

As one specific example of the generalizations of the Chaplygin gas, there is the modified Chaplygin gas \cite{Benaoum}-\cite{mcg2}, defined by the equation of state
\begin{equation}
p=\left(\gamma-1\right)\rho-M\rho^{-\mu}\,,
\end{equation}
where $p$ is the pressure of the fluid, $\rho$ its energy density, and $M$, $\mu$ and $\gamma$ are free parameters. Usually one restricts the studies of this gas to the range $\mu\geq -1$. If one chooses $\gamma=1$ one has the so-called generalized Chaplygin gas, while chossing $\gamma=\mu=1$ one has the original Chaplygin gas. 

From the condition for conservation of energy, with an adiabatic expansion of the universe quantified through the scale factor $a$, 
\begin{equation}
\frac{d\rho}{p+\rho}=-3\frac{da}{a}\,,
\end{equation}
one obtains, for $\gamma\neq 0$ and $\mu\neq -1$, the expression 
\begin{equation}
\rho=\left[A+\left(B-A\right)a^{-3\gamma\left(1+\mu\right)}\right]^{\frac{1}{1+\mu}}\,,
\end{equation}
where $A\equiv M/\gamma$ and $B\equiv \rho_0^{1+\mu}$. Such result, in conjunction with the Friedmann equation, 
\begin{equation}
H^2+\frac{k}{a^2}=\frac{8\pi}{3}\rho\,,
\end{equation}
where $H\equiv\dot{a}/a$, may yield solutions for $a=a\left(t\right)$.

For a flat space, Debnath, Banerjee and Chakraborty \cite{Debnath} give the general formula
\begin{equation}
\,_2F_1\left[x,x;1+x;-\frac{Aa^{3\gamma\left(1+\mu\right)}}{B-A}\right]=
a^{-\frac{3\gamma}{2}}t\sqrt{6\pi}\gamma \left(B-A\right)^{\frac{1}{2\left(1+\mu\right)}}\,,
\end{equation}
where $x\equiv 1/\left[2\left(1+\mu\right)\right]$, which can be inverted easily to yield $a=a\left(t\right)$ only for $\mu=0$ and $\mu=-1/2$. For spaces with curvature, explicit solutions for $a\left(t\right)$ for the particular choices $\mu=-1/2$, $\gamma=2/3$ and $\gamma=4/3$ are given by Costa \cite{BJP1}.
Therefore, instead of trying to obtain analytical solutions for $a\left(t\right)$ one can search for another useful variable which would play a role similar to the one of the cosmological time $t$. 

In the present work, the relations between the modified Chaplygin gas and a simple homogeneous cosmological scalar field are reevaluated, and the analytical results which are obtained are more general than the ones previously presented in the literature. Aside from a discussion about its physical implications (for example, the stability of the solutions found \cite{Perrotta}), it is simple to see that the scalar field representation of the modified Chaplygin gas can be a useful mathematical tool in cosmological studies, such as in the mathematical analysis of pertubations, since the scalar field can be pragmatically seen as a variable similar to the cosmological time.  

The structure of this work is the following: first the general formalism is presented, and in sequence two sections show analytic solutions for flat spaces and spaces with curvature. Finally, a last section presents some comments and conclusions about the results shown.

\section{The general formalism}
Both energy density and pressure of the modified Chaplygin gas can be related to a homogeneous scalar field $\varphi$ through the transformation equations 
\begin{equation}
\rho=\frac{\dot{\varphi}^2}{2}+V\left(\varphi\right)
\end{equation}
and
\begin{equation}
p=\frac{\dot{\varphi}^2}{2}-V\left(\varphi\right)\,,
\end{equation}
where the first term in the right side of each equality corresponds to the kinetical energy of the field, while the second one corresponds to its potential energial.

If one assumes that the value of field decreases with the expansion of the universe, one may write 
\begin{equation}
\dot{\varphi}=-\left(p+\rho\right)^{\frac12}\,.
\end{equation}
However,
\begin{equation}
\dot{\varphi}=\frac{d\varphi}{dt}=\frac{d\varphi}{da}\frac{da}{dt}=\frac{d\varphi}{da}aH\,.
\end{equation}
Finally, the Friedmann equation may be susbstituted in this last result to give
\begin{equation}
d\varphi=-da\left[\frac{p\left(a\right)+\rho\left(a\right)}{\frac{8\pi}{3}\rho\left(a\right)a^2-k}\right]^{\frac12}\,.
\end{equation}
It is not hard to notice that analytical solutions are more easily obtained when $k=0$. Anyway, the important fact to notice here is that one does not need to obtain an explicit solution for $a=a\left(t\right)$ in order to obtain $a=a\left(\varphi\right)$. This means that $\varphi$ may be seen as a surrogate quantity to be used in the place of the cosmological time $t$. 

\section{\label{flat}Flat spaces}
When the curvature parameter is null, one has
\begin{equation}
d\varphi=-\frac{da}{a}\sqrt{\frac{3}{8\pi}}\left[1+\frac{p\left(a\right)}{\rho\left(a\right)}\right]^{\frac12}\,,
\end{equation}
what, for the modified Chaplygin gas, becomes 
\begin{eqnarray}
d\varphi &=&-\frac{da}{a}\sqrt{\frac{3}{8\pi}}\left[1+\left(\gamma-1\right)-M\rho^{-1-\mu}\right]^{\frac12}\nonumber\\
&=&-\frac{da}{a}\sqrt{\frac{3\gamma}{8\pi}}\left[\frac{\left(B-A\right)a^{-3\gamma\left(1+\mu\right)}}{A+\left(B-A\right)a^{-3\gamma\left(1+\mu\right)}}\right]^{\frac12}\,.
\end{eqnarray}
For convenience one must assume from this point that $B-A>0$. However, the parameters $M$, $\mu$ and $\gamma$ may assume any real value. Now, the substitution
\begin{equation}
a^{-3\gamma\left(1+\mu\right)}=\frac{\left|A\right|\cosh 2u-A}{2\left(B-A\right)}\,,
\end{equation}
valid for $\gamma\neq 0$ and $\mu\neq -1$, yields
\begin{equation} 
\frac{da}{a}=\frac{-2du}{3\gamma\left(1+\mu\right)}\frac{\left|A\right|\sinh 2u}{\left|A\right|\cosh 2u-A}\,,
\end{equation}
and so
\begin{equation}
d\varphi=\frac{du}{\sqrt{6\pi\gamma}\left(1+\mu\right)}\,.
\end{equation}

\begin{figure}
\centerline{\includegraphics[scale=.45]{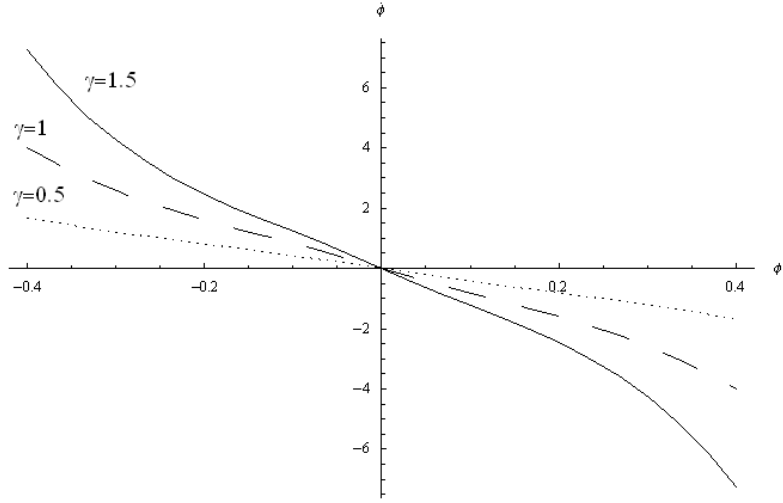}}
\caption{\label{graphdotvarphi}Graphs for $\dot{\phi}$ versus $\phi$, where $\phi=\varphi-\varphi_0$. In all graphs $A>0$ and $\mu=1$, while $\gamma$ assumes three values, 1/2, 1 and 3/2.}
\end{figure}

Therefore,
\begin{equation}
a=\left\{\frac{\left|A\right|\cosh\left[2\sqrt{6\pi\gamma}\left(1+\mu\right)\left(\varphi-\varphi_0\right)\right]-A}{2\left(B-A\right)}\right\}^{-\frac{1}{3\gamma\left(1+\mu\right)}}\,,
\end{equation}
with $\varphi_0$ being a constant of integration, and, consequently,
\begin{equation}
\rho=\left\{\frac{\left|A\right|}{2}\cosh\left[2\sqrt{6\pi\gamma}\left(1+\mu\right)\left(\varphi-\varphi_0\right)\right]+\frac{A}{2}\right\}^{\frac{1}{1+\mu}}\,,
\end{equation}
and
\begin{equation}
p=\rho\left\{\gamma\frac{\left|A\right|\cosh\left[2\sqrt{6\pi\gamma}\left(1+\mu\right)\left(\varphi-\varphi_0\right)\right]-A}{\left|A\right|\cosh\left[2\sqrt{6\pi\gamma}\left(1+\mu\right)\left(\varphi-\varphi_0\right)\right]+A}-1\right\}\,.
\end{equation}
With these one can obtain
\begin{equation}
\dot{\varphi}=-\left(\rho\gamma\right)^{\frac{1}{2}}\left\{\frac{\left|A\right|\cosh\left[2\sqrt{6\pi\gamma}\left(1+\mu\right)\left(\varphi-\varphi_0\right)\right]-A}{\left|A\right|\cosh\left[2\sqrt{6\pi\gamma}\left(1+\mu\right)\left(\varphi-\varphi_0\right)\right]+A}\right\}^{\frac12}
\end{equation}
and
\begin{equation}
V\left(\varphi\right)=\frac12\rho\left\{2-\gamma\frac{\left|A\right|\cosh\left[2\sqrt{6\pi\gamma}\left(1+\mu\right)\left(\varphi-\varphi_0\right)\right]-A}{\left|A\right|\cosh\left[2\sqrt{6\pi\gamma}\left(1+\mu\right)\left(\varphi-\varphi_0\right)\right]+A}\right\}\,.
\label{potential}
\end{equation}
Figure \ref{graphdotvarphi} shows the relation between $\dot{\varphi}$ and $\varphi$ for some values of $A>0$, $\mu$ and $\gamma$, where one can notice that all curves pass through the point $\left(\varphi_0,0\right)$. 

Substituting, in the equation of the modified Chaplygin gas, the values $\mu=\gamma=1$ one has the equation of state defining the original Chaplygin gas. Making these same substitutions in equation (\ref{potential}) and considering $A>0$, one has
\begin{equation}
V\left(\varphi\right)=\frac12\rho\left\{2-\frac{\cosh\left[4\sqrt{6\pi}\left(\varphi-\varphi_0\right)\right]-1}{\cosh\left[4\sqrt{6\pi}\left(\varphi-\varphi_0\right)\right]+1}\right\}\,,
\label{potential2}
\end{equation}
or, simply,
\begin{equation}
V\left(\varphi\right)=\frac12\sqrt{A}\left(\cosh 3\hat{\phi}+\frac{1}{\cosh 3\hat{\phi}}\right)\,,
\end{equation}
where $3\hat{\phi}\equiv 2\sqrt{6\pi}\left(\varphi-\varphi_0\right)$. This result is well known and appears in several places in the literature \cite{alternative}-\cite{Perrotta}. 

For the specific case of $A>0$, one can write
\begin{equation}
V\left(\varphi\right)=\frac{2-\gamma}{2}A^{\frac{1}{1+\mu}}\cosh^{\frac{2}{1+\mu}}\bar{\phi}+\frac{\gamma}{2}A^{\frac{1}{1+\mu}}\cosh^{-\frac{2\mu}{1+\mu}}\bar{\phi}\,,
\end{equation}
where $\bar{\phi}\equiv\left[\sqrt{6\pi\gamma}\left(1+\mu\right)\left(\varphi-\varphi_0\right)\right]$. This expression, valid only for $A>0$, can also be found in the literature \cite{Debnath,MakHarko}.

\begin{figure}
\centerline{\includegraphics[scale=0.45]{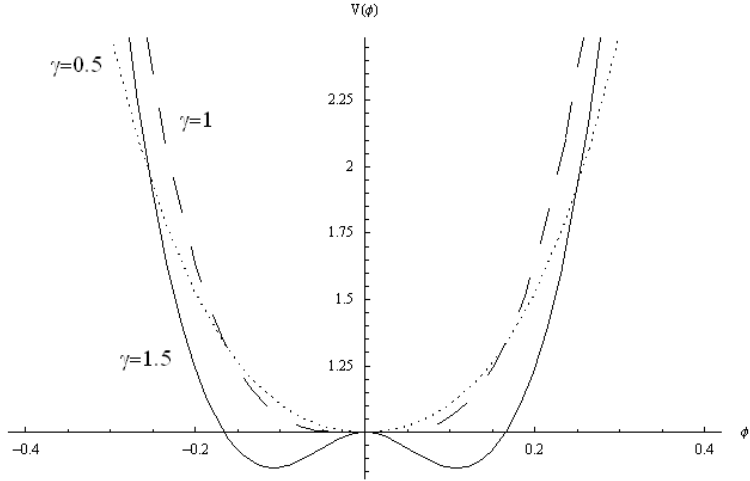}}
\caption{\label{graphpotvarphi}Graphs for the potential $V\left(\phi\right)$, where $\phi\equiv\varphi-\varphi_0$. In all graphs $A>0$ and $\mu=1$, while $\gamma$ assumes three values, 1/2, 1 and 3/2, and only for this last one the potential presents two minima.}
\end{figure}

One interesting feature of the potential associated to the modified Chaplygin gas, the possible presence of more than one minimum, is not emphasized explicitly anywhere in previous studies\footnote{To be fair, an attentive reader can notice, both in the text and in the graphs, that one of the potentials presented by Debnath, Banerjee and Chakraborty \cite{Debnath} has two minima.}. However, to see this, one needs only to analyze the derivative of the potential of the scalar field, 
\begin{equation}
\frac{\partial V}{\partial \varphi}=\frac12\frac{\partial\rho}{\partial\varphi}\left(2-\gamma-\mu M\rho^{-\mu-1}\right)\,,
\end{equation}
which, when null, indicates the possible minima or maxima of the potential, i.e.,
\begin{equation}
\frac{\partial V}{\partial \varphi}=0\Rightarrow\left\{
\begin{array}{l}
\varphi=\varphi_0\\
\varphi=\varphi_0+\frac{1}{2\sqrt{6\pi\gamma}\left(1+\mu\right)}\;\mathrm{arccosh}\;\left[\frac{\gamma\left(2\mu+1\right)-2}{2-\gamma}\frac{A}{\left|A\right|}\right]
\end{array}
\right.\,.
\end{equation}
The last condition is obeyed if
\begin{equation}
\label{condition1}
\frac{\gamma\left(2\mu+1\right)-2}{2-\gamma}\frac{A}{\left|A\right|}\geq 1\,.
\end{equation}
The case $\gamma=2$ is a particular one, and it presents only a maximum at $\varphi=\varphi_0$. Figure \ref{graphpotvarphi} shows the potential $V\left(\varphi\right)$ for some values of $A>0$, $\mu$ and $\gamma$, showing clearly a case where the potential presents two minima.

\section{Spaces with curvature}
If the curvature parameter, $k$, is not zero, it is not easy to obtain analytic solutions for the functions $a\left(\varphi\right)$ and $V\left(\varphi\right)$, and finding such solutions seems to be a task feasible only for certain values of the parameters $\mu$ and $\gamma$. 

A first example which allows an analytic solution is obtained when $\gamma=4/3$ and $\mu=0$. In such case one has
\begin{equation}
\rho=A+\left(B-A\right)a^{-4}\,,
\end{equation}
and, therefore, one has an equivalence to a universe filled with radiation and a cosmological constant.
The analytic solution for the scale factor $a\left(\varphi\right)$ is
\begin{equation}
a=\frac{\left[\frac{4\left(B-A\right)}{\left|A\right|}\right]^{\frac14}e^{\sqrt{2\pi}\left(\varphi-\varphi_0\right)}}{\left\{\left[e^{\sqrt{8\pi}\left(\varphi-\varphi_0\right)}+\frac{3k}{16\pi\sqrt{\left|A\right|\left(B-A\right)}}\right]^2-\frac{A}{\left|A\right|}\right\}^{\frac12}}\,,
\end{equation}
and it is possible to obtain also $a\left(t\right)$ and $\varphi\left(t\right)$.

Since, now,
\begin{equation}
V\left(\varphi\right)=A+\frac{B-A}{3a^4}\,,
\end{equation}
one has
\begin{equation}
\frac{\partial V}{\partial\varphi}=-\frac{4\left(B-A\right)}{3a^5}\frac{\partial a}{\partial\varphi}\,,
\end{equation}
and the potential may have two minima given by the conditions
\begin{equation}
\exp\left[\sqrt{8\pi}\left(\varphi_m-\varphi_0\right)\right]+\frac{3k}{16\pi\sqrt{\left|A\right|\left(B-A\right)}}=\pm\sqrt{\frac{A}{\left|A\right|}}
\end{equation}
and 
\begin{equation}
\exp\left[2\sqrt{8\pi}\left(\varphi_m-\varphi_0\right)\right]=\frac{9k^2}{256\pi^2\left|A\right|\left(B-A\right)}-\frac{A}{\left|A\right|}\,.
\end{equation}
Clearly, the first one is valid only if $A>0$, while the second one may occur only for $k\neq 0$ if $A>0$. As a matter of fact, it is important to notice that in the flat case only one one of the above conditions may be obeyed, and only in the presence of curvature the two may be valid simultaneously.

As another example involving curvature, not easily solvable in terms of the cosmological time $t$, and not found previously in the literature, there is the case where $\mu=-2$ and $\gamma=2/3$, when
\begin{equation}
\rho=\left[A+\left(B-A\right)a^{2}\right]^{-1}\,.
\end{equation}
An important feature of this case, which is not very physically appealing, is that the signal of the acceleration of the universe depends solely on the signal of $A$, i.e.,
\begin{equation}
\frac{\ddot{a}}{a}=\frac{8\pi}{3}A\rho^2\,.
\end{equation}

To obtain the explicit analytical solutions one begins with 
\begin{eqnarray}
p+\rho &=&\frac{2}{3}\rho-M\rho^2\nonumber\\
&=&\frac{2\left(B-A\right)a^{2}}{3\left[A+\left(B-A\right)a^{2}\right]^2}
\end{eqnarray}
and, therefore, one has
\begin{eqnarray}
d\varphi 
&=&-\frac{\left[2\left(B-A\right)\right]^{\frac12}ada}{\left[A+\left(B-A\right)a^{2}\right]^{\frac12}\left\{8\pi a^2-3k\left[A+\left(B-A\right)a^{2}\right]\right\}^{\frac12}}\nonumber\\
&=&-\frac{dz}{\left\{\left[4\pi-\frac32 k\left(B-A\right)\right]z^2-4\pi A\right\}^\frac12}\,,
\end{eqnarray}
where $z\equiv\left[A+\left(B-A\right)a^2\right]^\frac12$.
This last result easily yields
\begin{equation}
a^2=\frac{\left|A\right|\cosh\left[-2\sqrt{4\pi-\frac{3k\left(B-A\right)}{2}}\left(\varphi-\varphi_0\right)\right]-A\left[1-\frac{3k\left(B-A\right)}{4\pi}\right]}{2\left(B-A\right)\left[1-\frac{3k\left(B-A\right)}{8\pi}\right]}
\label{a2}
\end{equation}
and, thus, one has
\begin{equation}
\rho=\frac{1-\frac{3k}{8\pi}\left(B-A\right)}{\frac12\left\{\left|A\right|\cosh\left[-2\sqrt{4\pi-\frac{3k\left(B-A\right)}{2} }\left(\varphi-\varphi_0\right)\right]+A\right\}}\,,
\end{equation}
\begin{eqnarray}
p&=&-\frac\rho 3\left(1+2A\rho\right)\nonumber\\
&=&-\frac\rho 3\left\{1+\frac{4\frac{A}{\left|A\right|}\left[1-\frac{3k}{8\pi}\left(B-A\right)\right]}{\cosh\left[-2\sqrt{4\pi-\frac{3k\left(B-A\right)}{2}}\left(\varphi-\varphi_0\right)\right]+\frac{A}{\left|A\right|}}\right\}\,,
\end{eqnarray}
and, finally, the potential
\begin{equation}
V\left(\varphi\right)=\frac\rho 3\left\{2+\frac{2\frac{A}{\left|A\right|}\left[1-\frac{3k}{8\pi}\left(B-A\right)\right]}{\cosh\left[-2\sqrt{4\pi-\frac{3k\left(B-A\right)}{2}}\left(\varphi-\varphi_0\right)\right]+\frac{A}{\left|A\right|}}\right\}\,.
\end{equation}

Since, in this case, 
\begin{equation}
\frac{\partial V}{\partial \varphi}=\frac{2}{3}\frac{\partial \rho}{\partial\varphi}\left(1+A\rho\right)\,,
\end{equation}
the potential $V\left(\varphi\right)$ will have two minima only if $\rho=-1/A$, and for a physically relevant $\rho>0$, this is obtained only if $A<0$, when the acceleration of the universe is negative. Anyway, it is interesting to look at the condition for the occurrence of the second minimum, explicitly, which is 
\begin{equation}
\frac{\cosh\left[-2\sqrt{4\pi-\frac{3k\left(B-A\right)}{2}}\left(\varphi_{m}-\varphi_0\right)\right]}{k\left(B-A\right)-4\pi}=\frac{3A}{4\pi\left|A\right|}\,,
\end{equation}
where $\varphi_{m}$ is the point of possible occurrence of the minimum. One can notice that if $k>0$ one could have this condition fullfilled even for $A>0$, but at the price of having $\rho<0$.

\section{Final comments}
The relations between the modified Chaplygin gas and a cosmological scalar field shown in this work indicate that models using the modified Chaplygin gas as a single fluid may also be studied using a representation in terms of a scalar field. For example, in cosmology one may be interested in solutions for the scale factor when there is curvature. Since, for cosmologies with the modified Chaplygin gas acting as a single fluid, few of such analytic solutions are known, the usage of a scalar field as an auxiliary quantity offers, at least in principle, another way for the search of new solutions.    

Also, since mathematically is possible to have $A<0$, it is at least interesting to have solutions including such possibility. In the literature, these solutions, if previously existent, are not broadly discussed. Here, all the solutions shown are valid for any value of $A$ satisfying the condition $A<\rho_0^{1+\mu}$. It is interesting to notice that the conjunction between solutions with several values of curvature and $A$ may yield novelties in the potential of the field, $V\left(\varphi\right)$, which may present more than one point of minimum.

Finally, the scalar field representation may also be of some utility, for example, in the mathematical analysis of the evolution of perturbations \cite{GRG}, where the relevant quantity, the density contrast $\delta$, is usually seen as a function of the cosmological time $t$, the conformal time $\eta$ or the scale factor $a$. To give an specific example, using the results of Section \ref{flat}, and without considering the ``averaging problem'' \cite{linear}, one can rewrite the equation for the perturbations (cf. equation 4.122 from Padmanabhan \cite{Padmanabhan}),
\begin{equation}
\frac{d^2\delta}{da^2}+\frac{3-15\omega+6v^2}{2a}\frac{d\delta}{da}+\frac{k^2v^2\delta}{H^2a^4}=\frac{3\delta}{2a^2}\left(1-6v^2-3\omega^2+8\omega\right)\,,
\end{equation}
using as variable
\begin{equation}
w=\frac{1}{2}\left(1-\frac{\left|A\right|}{A}\cosh 2u\right)=-\left(\frac{B-A}{A}\right)a^{-3\gamma\left(1+\mu\right)}\,,
\end{equation}
where 
\begin{equation}
u\equiv\sqrt{6\pi\gamma}\left(1+\mu\right)\left(\varphi-\varphi_0\right)\,,
\end{equation}
and, by doing this, one can obtain a very general solution for the wavemode $k=0$, 
\begin{equation}
\delta=\frac{w^{\frac{2x}{\gamma}}}{\left(1-w\right)^x}\left\{c_1+c_2w^{1-\frac{2x}{3\gamma}}\,_2F_1\left[1-\frac{2x}{3\gamma},1+x;2-\frac{2x}{3\gamma};w\right]\right\}\,,
\end{equation}
where $x\equiv1/\left[2\left(1+\mu\right)\right]$, with $c_1$ and $c_2$ being arbitrary constants. It is important to notice that this is a new result which incorporates the possibility of having a negative value for $A$. Therefore, the representation of the modified Chaplygin gas in terms of a scalar field $\varphi$ opens another road for the study of the evolution of perturbations, and as such it may be seen as a mathematical tool of some value. 


\end{document}